\documentclass[12pt,english]{article}
\usepackage[T1]{fontenc}
\usepackage[latin9]{inputenc}
\usepackage{geometry}
\usepackage{color}
\usepackage{float}
\usepackage{amsmath}
\usepackage{amssymb}
\usepackage{setspace}
\usepackage{wasysym}
\usepackage[authoryear]{natbib}
\makeatletter

\providecommand{\tabularnewline}{\\}

\newtheorem{assumption}{Assumption}

\newcommand*{\indep}{%
  \mathbin{%
    \mathpalette{\@indep}{}%
  }%
}
\newcommand*{\nindep}{%
  \mathbin{
    \mathpalette{\@indep}{\not}
  }%
}
\newcommand*{\@indep}[2]{%
  \sbox0{$#1\perp\m@th$}
  \sbox2{$#1=$}
  \sbox4{$#1\vcenter{}$}
  \rlap{\copy0}
  \dimen@=\dimexpr\HT2-\HT4-.2pt\relax
  \kern\dimen@
  {#2}%
  \kern\dimen@
  \copy0 
} 

\newcommand{\pr}{P} 
\newcommand{\var}{\mathrm{var}}
\newcommand{\HT}{\mathrm{HT}}
\newcommand{\cov}{\mathrm{cov}}

\newcommand{\obs}{\mathrm{obs}}

\newcommand{\R}{\mathbb{R}}
\newcommand{\de}{\mathrm{d}}
\newcommand{\T}{\mathrm{\scriptscriptstyle T}}

\newcommand{\ipw}{\mathrm{IPW}}

\newcommand{\gls}{\mathrm{GLS}}
\newcommand{\Cal}{\mathrm{cal}}
\newcommand{\nni}{\mathrm{nni}}

\newcommand{\dr}{\mathrm{dr}}

\newcommand{\N}{\mathcal{N}}
\newcommand{\U}{\mathcal{U}}
\newcommand{\F}{\mathcal{F}}

\newcommand{\I}{\mathrm{I}}

\usepackage{babel}

\usepackage{babel}

\usepackage{babel}

\usepackage{babel}

\makeatother

\usepackage{babel}
\begin{document}
\title{\textbf{Statistical Data Integration in Survey Sampling: A review}}
\author{Shu Yang \thanks{Department of Statistics, North Carolina State University}
\and Jae Kwang Kim \thanks{Department of Statistics, Iowa State University} }
\date{\today}

\maketitle
 
\begin{abstract}
Finite population inference is a central goal in survey sampling.
Probability sampling is the main statistical approach to finite population
inference. Challenges arise due to high cost and increasing non-response
rates. Data integration provides a timely solution by leveraging multiple
data sources to provide more robust and efficient inference than using
any single data source alone. The technique for data integration varies
depending on types of samples and available information to be combined.
This article provides a systematic review of data integration techniques
for combining probability samples,  probability and non-probability
samples, and probability and big data samples. We discuss
a wide range of integration methods such as generalized least squares,
calibration weighting, inverse probability weighting, mass imputation
and doubly robust methods. Finally, we highlight important questions
for future research.

\textit{key words}: Data fusion; Generalizability; Meta analysis;
Missingness at random; Transportability 
\end{abstract}
\newpage{}

\section{Introduction}

Probability sampling is regarded as the gold-standard in survey statistics
for finite population inference. Fundamentally, probability samples
are selected under known sampling designs and therefore are representative
of the target population. Because the selection probability is known,
the subsequent inference from a probability sample is often design-based
and respects  the way in which the data were collected; see \citet{sarndal2003model,cochran2007sampling,fuller2009sampling}
for textbook discussions. \citet{kalton2019} provided a comprehensive
overview of the survey sampling research in the last 60 years.

However, many practical challenges arise in collecting and analyzing
probability sample data (\citealp{baker13}; \citealp{keiding2016perils}).
Large-scale survey programs continually face heightened demands coupled
with reduced resources. Demands include requests for estimates for
domains with small sample sizes and desires for more timely estimates.
Simultaneously, program budget cuts force reductions in sample sizes,
and decreasing response rates make non-response bias an important
concern.

Data integration is a new area of research to provide a timely solution
to the above challenges. The goal is multi-fold: i) minimize the cost
associated with surveys, ii) minimize the respondent burden and iii)
maximize the statistical information or equivalently the efficiency
of survey estimation. Narrowly speaking, survey integration means
combining separate probability samples into one survey vehicle \citep{bycroft2010integrated}.
Broadly speaking, one can consider combining probability samples with
non-probability samples. Recently in survey statistics, non-probability
data become increasingly available for research purposes and provide
unprecedented opportunities for new scientific discovery; however,
they also present \textcolor{black}{additional} challenges such as
heterogeneity, selection bias, high dimensionality, etc. The past
years have seen immense progress in theories, methods, and algorithms
for surmounting important challenges arising from non-probability
data analysis. This article provides a systematic review of data integration
for combining probability samples, probability and non-probability
samples, and probability and big data samples.

Section \ref{sec:integProbSamples} establishes notation and reviews
these methods \textcolor{black}{in the context of combining multiple
probability samples}. Existing methods for probability data integration
can be categorized into two types depending on the level of information
to be combined: a macro approach combining the summary statistics
from multiple surveys and a micro approach creating synthetic imputations.

Section \ref{sec:PNP} describes the motivation, challenges and methods
for integrating probability and emergent non-probability samples.
We also draw connections of survey data integration to combine randomized
clinical trials and real-world data in Biostatistics. We then  discuss a
wide range of integration methods including calibration weighting,
inverse probability weighting, mass imputation and doubly robust methods.

We then consider data integration methods for combining probability
and big non-probability samples. Depending on the roles in statistical
inference, there are two types of \textit{big data}: one with large
sample sizes (large $n$) and the other with rich covariates (large
$p$). In the first type, the non-probability sample can be large
in sample size. How to leverage the rich information in the big data
to improve the finite population inference is an important research.
In the second type, there are a large number of variables. There is
a large literature on variable selection methods for prediction, but
little work on variable selection for data integration that \textcolor{black}{can
successfully recognize the strengths and the limitations of each data
source and utilize all information captured for finite population
inference.} Section \ref{sec:PbigD} presents robust data integration
and variable selection methods in this context.

To summarize, Section \ref{sec:Concluding-remark} describes the direction
of future research along the line of data integration including sensitivity
analysis to assess the robustness of study conclusions to unverifiable
assumptions, hierarchical modeling, and some cautionary remarks.

\section{Combining probability samples\label{sec:integProbSamples}}

\subsection{Multiple probability samples and missingness patterns}

Combining two or more independent survey probability samples is a
problem frequently encountered in \textcolor{black}{the practice of
} survey sampling. For simplicity of exposition, let $\U=\{1,\ldots,N\}$
be the index set of $N$ units for the finite population, with $N$
being the known population size. Let $(x_{i}^{\T},y_{i})^{\T}$ be
the realized value of a vector of random variables $(X^{\T},Y)^{\T}$
for unit $i$. Let $I_{i}$ be the sample indicator such that $I_{i}=1$
indicates the selection of unit $i$ into the sample and $I_{i}=0$
otherwise. The probability $\pi_{i}=\pr(I_{i}=1\mid i\in\U)$ is called
the first-order inclusion probability and is known by \textcolor{black}{the
sampling design}. The design weight is $d_{i}=\pi_{i}^{-1}$. The
joint probability $\pi_{ij}=\pr(I_{i}I_{j}=1\mid i,j\in\U)$ is called
the second-order inclusion probability and is often used for variance
estimation of the design-weighted estimator. \textcolor{black}{The
sample size is} $n=\sum_{i=1}^{N}I_{i}.$

The \textcolor{black}{main} advantage of probability sampling is to
ensure design-based inference. For example, the Horvitz-Thompson (HT)
estimator of the population mean of $y$, denoted by $\mu_{y}$, is
$\widehat{\mu}_{\HT}=N^{-1}\sum_{i:I_{i}=1}\pi_{i}^{-1}y_{i}$, and
the design-variance estimator is

\[
\widehat{V}_{\HT}=nN^{-2}\sum_{i:I_{i}=1}\sum_{j:I_{j}=1}\frac{(\pi_{ij}-\pi_{i}\pi_{j})}{\pi_{ij}}\frac{y_{i}}{\pi_{i}}\frac{y_{j}}{\pi_{j}}.
\]
We consider multiple sources of probability data. For multiple datasets,
we use the subscript letter to indicate the respective sample; for
example, we use $d_{A,i}$ as the design weight of unit $i$ in sample
A.

Depending on the available information from multiple data sources,
\textcolor{black}{each sample has planned missingness by design.}
As illustrated in Table \ref{tab:Missingness-patterns}, the combined
sample exhibits different missingness patterns: monotone and non-monotone.
For monotone missingness, our framework covers two common types of
studies. First, we have a large main dataset, and then collect more
information on important variables for a subset of units, e.g., using
a two-phase sampling design \citep{neyman1938contribution,cochran2007sampling,wang2009causal}.
Consider the U.S. Census of housing and population as an example.
The short form consists of $100\%$ sample, for which basic demographic
information was obtained. The long form consists about $16\%$ sample,
for which other social and economic information as well as demographic
information were obtained. \citet{deming1940least} considered this
setup as a classical two-phase sampling problem and use calibration
weighting for demographic variable to match the known population counts
from the short form.

Second, we have a smaller and carefully designed validation dataset
with rich covariates, and then link it to a larger main dataset with
fewer covariates. The setup of two independent samples with common
items is often called non-nested two-phase sampling. Consider the
U.S. consumer expenditure survey as an example. Two independent samples
were selected from the same finite population, including a diary survey
sample, referred to as sample A, and a face-to-face survey sample,
referred to as sample B. In sample A, observe auxiliary information
$X$ and outcome $Y$; whereas in sample B, observe common auxiliary
information $X$. \citet{zieschang1990sample} considered using sample
weighting to estimate \textcolor{black}{{} detailed expenditure and
income items} combining sample A and sample B. Another example is
the Canadian Survey of Employment, Payrolls and Hours considered by
\citet{hidi01}. Sample A is a small sample from Statistics Canada
Business Register, in which the study variables $Y$, number of hours
worked by employees and summarized earnings, were observed. Sample
B is a large sample drawn from a Canadian Customs and Revenue Agency
administrative data, in which auxiliary variables $X$ were observed.

Finally, we will consider combining two independent surveys with non-monotone missing patterns. Statistical matching technique will be introduced in Section 2.3 as a general statistical tool under this setup.  

\begin{table}[H]
\caption{\label{tab:Missingness-patterns}Missingness patterns in the combined
samples: ``$\checked$'' means ``is measured''}

\centering%
\begin{tabular}{ccccc}
\hline 
\multicolumn{5}{c}{Monotone missingness}\tabularnewline
\hline 
 & $d$  & $X$  & $Y$  & \tabularnewline
Sample A  & $\checked$  & $\checked$  & $\checked$  & \tabularnewline
Sample B  & $\checked$  & $\checked$  &  & \tabularnewline
\hline 
\hline 
\multicolumn{5}{c}{Non-monotone missingness I}\tabularnewline
\hline 
 & $d$  & $X$  & $Y_{1}$  & $Y_{2}$\tabularnewline
Sample A  & $\checked$  & $\checked$  & $\checked$  & $\checked$\tabularnewline
Sample B  & $\checked$  & $\checked$  & $\checked$  & \tabularnewline
Sample C  & $\checked$  & $\checked$  &  & $\checked$\tabularnewline
\hline 
\hline 
\multicolumn{5}{c}{Non-monotone missingness II}\tabularnewline
\hline 
 & $d$  & $X$  & $Y_{1}$  & $Y_{2}$\tabularnewline
Sample A  & $\checked$  & $\checked$  & $\checked$  & \tabularnewline
Sample B  & $\checked$  & $\checked$  &  & $\checked$\tabularnewline
\hline 
\end{tabular}

$d$ is the sampling weight, where the subscript indicates the sample,
$X$ is the vector of auxiliary variables, $Y$, $Y_{1}$ and $Y_{2}$
are scalar outcome variables. 
\end{table}

\subsection{Two approaches for probability data integration}

We classify probability data integration methods based on the level
of information to be combined: a macro approach and a micro approach.
In the macro approach, we obtain summary information such as the point
and variance estimates from multiple data sources and combine those
\textcolor{black}{to obtain} a more efficient estimator of the parameter
of interest. In the micro approach, we create a single synthetic data
that contains all available information from all data sources.

\subsubsection{Macro approach: Generalized least squares (GLS) estimation}

\citet{ren97}, \citet{hidi01}, \citet{mer04}, \citet{wu04}, \citet{ybarra08}
and \citet{mer10} considered the problem of combining data from two
independent probability samples to estimate totals at the population
and domain levels. \citet{mer04} and \citet{mer10} provided a rigorous
treatment of the survey integration through the generalized method
of moments.

We focus on the monotone missingness pattern. The same discussion
applies to the other patterns. From each probability sample, we obtain
different estimators for the means of common items. The GLS approach
combines those estimates as an optimal estimator. Let $\widehat{\mu}_{x,A}$
and $\widehat{\mu}_{x,B}$ be unbiased estimators of $\mu_{x}$ from
sample A and sample B, respectively. Let $\widehat{\mu}_{B}$ be an
unbiased estimator of $\mu_{y}$ from sample B.

To combine the multiple estimates, \textcolor{black}{we may build
a linear model of three estimates with two parameters as follows:}
\begin{equation}
\left(\begin{array}{c}
\widehat{\mu}_{x,A}\\
\widehat{\mu}_{x,B}\\
\widehat{\mu}_{B}
\end{array}\right)=\left(\begin{array}{cc}
1 & 0\\
1 & 0\\
0 & 1
\end{array}\right)\left(\begin{array}{c}
\mu_{x}\\
\mu_{y}
\end{array}\right)+\left(\begin{array}{c}
e_{1}\\
e_{2}\\
e_{3}
\end{array}\right),\label{eq:GLS}
\end{equation}
where $(e_{1},e_{2},e_{3})^{\T}$ has mean $(0,0,0)^{\T}$, variance-covariance
\[
V=\left(\begin{array}{ccc}
\var(\widehat{\mu}_{x,A}) & \cov(\widehat{\mu}_{x,A},\widehat{\mu}_{x,B}) & \cov(\widehat{\mu}_{x,A},\widehat{\mu}_{B})\\
\cov(\widehat{\mu}_{x,A},\widehat{\mu}_{x,B}) & \var(\widehat{\mu}_{x,B}) & \cov(\widehat{\mu}_{x,B},\widehat{\mu}_{B})\\
\cov(\widehat{\mu}_{x,A},\widehat{\mu}_{B}) & \cov(\widehat{\mu}_{x,B},\widehat{\mu}_{B}) & \var(\widehat{\mu}_{B})
\end{array}\right),
\]
and $\var(\cdot)$ and $\cov(\cdot)$ are the variance and covariance
induced by the sampling probability.

Based on model (\ref{eq:GLS}), treat $(\widehat{\mu}_{x,A},\widehat{\mu}_{x,B},\widehat{\mu}_{B})$
as observations and define a sum of squared error term 
\[
Q(\mu_{x},\mu_{y})=\left(\begin{array}{c}
\widehat{\mu}_{x,A}-\mu_{x}\\
\widehat{\mu}_{x,B}-\mu_{x}\\
\widehat{\mu}_{B}-\mu_{y}
\end{array}\right)^{\T}V^{-1}\left(\begin{array}{c}
\widehat{\mu}_{x,A}-\mu_{x}\\
\widehat{\mu}_{x,B}-\mu_{x}\\
\widehat{\mu}_{B}-\mu_{y}
\end{array}\right).
\]
The optimal estimator of $(\mu_{x},\mu_{y})$ that minimizes $Q(\mu_{x},\mu_{y})$
is 
\begin{equation}
\widehat{\mu}_{x}^{*}=\alpha^{*}\widehat{\mu}_{x,A}+(1-\alpha^{*})\widehat{\mu}_{x,B}\label{eq:(3)}
\end{equation}
and 
\begin{equation}
\widehat{\mu}_{\gls}=\widehat{\mu}_{B}+\left(\begin{array}{c}
\widehat{\cov}(\widehat{\mu}_{x,A},\widehat{\mu}_{B})\\
\widehat{\cov}(\widehat{\mu}_{x,B},\widehat{\mu}_{B})
\end{array}\right)^{\T}\left(\begin{array}{cc}
\widehat{\var}(\widehat{\mu}_{x,A}) & \widehat{\cov}(\widehat{\mu}_{x,A},\widehat{\mu}_{x,B})\\
\widehat{\cov}(\widehat{\mu}_{x,A},\widehat{\mu}_{x,B}) & \widehat{\var}(\widehat{\mu}_{x,B})
\end{array}\right)^{-1}\left(\begin{array}{c}
\widehat{\mu}_{x}^{*}-\widehat{\mu}_{x,A}\\
\widehat{\mu}_{x}^{*}-\widehat{\mu}_{x,B}
\end{array}\right),\label{eq:(4)}
\end{equation}
where 
\[
\alpha^{*}=\frac{\widehat{\var}(\widehat{\mu}_{x,B})-\widehat{\cov}(\widehat{\mu}_{x,A},\widehat{\mu}_{x,B})}{\widehat{\var}(\widehat{\mu}_{x,A})+\widehat{\var}(\widehat{\mu}_{x,B})-2\widehat{\cov}(\widehat{\mu}_{x,A},\widehat{\mu}_{x,B})}.
\]

To see the efficiency gain of $\widehat{\mu}_{\gls}$ over $\widehat{\mu}_{B}$,
using (\ref{eq:(3)}), we express 
\[
\widehat{\mu}_{\gls}=\widehat{\mu}_{B}-\widehat{\cov}(\widehat{\mu}_{B},\widehat{\mu}_{x,B}-\widehat{\mu}_{x,A})\left\{ \widehat{\var}(\widehat{\mu}_{x,B}-\widehat{\mu}_{x,A})\right\} ^{-1}(\widehat{\mu}_{x}^{*}-\widehat{\mu}_{x,B}).
\]
The variance of $\widehat{\mu}_{\gls}$ is 
\[
\var(\widehat{\mu}_{B})-\cov(\widehat{\mu}_{B},\widehat{\mu}_{x,B}-\widehat{\mu}_{x,A})\left\{ \var(\widehat{\mu}_{x,B}-\widehat{\mu}_{x,A})\right\} ^{-1}\cov(\widehat{\mu}_{B},\widehat{\mu}_{x,B}-\widehat{\mu}_{x,A}),
\]
which is not larger than $\var(\widehat{\mu}_{B})$. \textcolor{black}{The
GLS estimator for non-monotone missingness can be constructed similarly.
See \citet{fuller99} for an application in the National Resource Inventory. 
}


\subsubsection{Micro approach: mass imputation}

Mass imputation (also called synthetic data imputation) is a technique
of creating imputed values for items not observed in the current survey
by incorporating information from other surveys. \citet{breidt1996two}
discussed mass imputation for two-phase sampling. \citet{rivers2007}
proposed a mass imputation approach using nearest neighbor imputation
but the theory is not fully developed. \citet{schenker07} reported
several applications of synthetic data imputation, using a model-based
method to estimate totals and other parameters associated with variables
not observed in a larger survey but observed in a much smaller survey.
\citet{legg2009two} and \citet{kim2012combining} developed synthetic
imputation approaches to combining two surveys. \citet{chipperfield2012combining}
discussed composite estimation when one of the surveys is mass imputed.
\citet{bethlehem2016solving} discussed practical issues in sample
matching \textcolor{black}{for mass imputation}.

The primary goal is to create a single synthetic dataset of proxy
values $\widehat{y}_{i}$ for the unobserved $y_{i}$ in sample B
and then use the proxy data together with the associated design weights
of sample A to produce projection estimators of the population mean
$\mu_{y}$. This is particularly useful when sample B is a large scale
survey and item $Y$ is very expensive to measure. The proxy values
$\widehat{y}_{i}$ are generated by first fitting a working model
relating $Y$ to $X$, $E(Y\mid X)=m(X;\beta_{0})$ based on the data
$\{(x_{i},y_{i}):i\in A\}$ from sample A. Then the synthetic values
of $Y$ can be created by $\widehat{y}_{i}=m(x_{i};\widehat{\beta})$
for $i\in B$. \textcolor{black}{Thus, sample A is used as a training
sample for predicting $Y$ in sample B.} The mass imputation estimator
of $\mu_{y}$ is $\widehat{\mu}_{\I}=N^{-1}\sum_{i\in B}d_{B,i}\widehat{y}_{i}$.
\citet{kim2012combining} showed that $\widehat{\mu}_{\I}$ is asymptotically
design-unbiased if $\widehat{\beta}$ satisfies 
\begin{equation}
\sum_{i\in A}d_{A,i}\{y_{i}-m(x_{i};\widehat{\beta})\}=0.\label{eq:model-cal}
\end{equation}
With (\ref{eq:model-cal}), 
\begin{eqnarray*}
\widehat{\mu}_{\I} & = & N^{-1}\sum_{i\in B}d_{B,i}\widehat{y}_{i}+N^{-1}\sum_{i\in A}d_{A,i}(y_{i}-\widehat{y}_{i})\\
 & = & N^{-1}\sum_{i\in B}d_{B,i}m(x_{i};\beta_{0})+N^{-1}\sum_{i\in A}d_{A,i}\{y_{i}-m(x_{i};\beta_{0})\}=\widehat{P}_{B}+\widehat{Q}_{A},
\end{eqnarray*}
and 
\[
\var(\widehat{\mu}_{\I})=\var(\widehat{P}_{B})+\var(\widehat{Q}_{A}).
\]

The asymptotic unbiasedness holds regardless of whether the regression
model is true or not. However, a good regression model will reduce
the variance of $\widehat{\mu}_{\I}$. For variance estimation, either
linearization or replication-based sampling \citep{kim2012combining}
can be used.

\subsection{Mass imputation with non-monotone missingness}

For non-monotone missingness, \textcolor{black}{{} the mass imputation
method of \citet{kim2012combining} is not directly applicable as
the sample with partial observations may contain additional information
for parameter estimation.} Often, one can consider a joint model of
all variables and use the EM algorithm to estimate the model parameters.
The joint model deduces the conditional distribution of the missing
variables given the observed values for imputation.

For illustration, consider the non-monotone missingness I structure
in Table \ref{tab:Missingness-patterns}. The goal is to develop mass
imputation for both $Y_{2}$ in sample B and $Y_{1}$ in sample C.
It is attempting to specify the conditional distribution of $Y_{2}$
given $(X,Y_{1})$ to impute $Y_{2}$ in sample B and the conditional
distribution of $Y_{1}$ given $(X,Y_{2})$ to impute $Y_{1}$ in
sample C. However, this approach may result in model incompatiability.
That is, there does not exist a joint model of $(Y_{1},Y_{2})$ given
$X$ that leads to the corresponding conditional distributions. To
avoid model incompatibility, we use a joint model for $(Y_{1},Y_{2})$
given $X$ for prediction though specifying the sequential conditional
distribution 
\begin{equation}
f(Y_{1},Y_{2}\mid X;\theta)=f_{1}(Y_{1}\mid X;\theta_{1})f_{2}(Y_{2}\mid X,Y_{1};\theta_{2}),\label{eqn5}
\end{equation}
where $\theta=(\theta_{1}^{\T},\theta_{2}^{\T})^{\T}$, $\theta_{1}$
and $\theta_{2}$ are unknown parameters.

For parameter estimation, it suffices to use observations in sample
A; however, this approach ignores the partial information in sample
B and sample C and therefore is not efficient. Let the joint set of
sampling indexes be $S=A\cup B\cup C$. \textcolor{black}{{} Assuming
no overlap between the samples, we define
\[
\pi_{S,i}=\pr(i\in S\mid i\in\U)=\left\{ \begin{array}{ll}
\pi_{A,i} & \mbox{if }i\in A\\
\pi_{B,i} & \mbox{if }i\in B\\
\pi_{C,i} & \mbox{if }i\in C
\end{array}\right.
\]
}and let $d_{i}$ be the design weight for unit $i\in S$ without
specifying which sample it belongs to. That is,
$d_{i}=d_{A,i}$ if $i\in A$.  To incorporate all available information,
the EM algorithm can be used as follows. 
\begin{description}
\item [{{{[}E-step}{]}}] Let $\theta^{(t)}$ be the parameter estimate
at iteration $t.$ Compute the conditional expectation of the pseudo
log-likelihood functions: 
\begin{eqnarray*}
Q_{1}(\theta_{1}\mid\theta^{(t)}) & = & \sum_{i\in S}d_{i}E\left\{ \log f_{1}(y_{1i}\mid x_{i};\theta_{1})\mid x_{i},y_{i,\obs};\theta^{(t)}\right\} \\
Q_{2}(\theta_{2}\mid\theta^{(t)}) & = & \sum_{i\in S}d_{i}E\left\{ \log f_{2}(y_{2i}\mid x_{i},y_{1i};\theta_{2})\mid x_{i},y_{i,\obs};\theta^{(t)}\right\} 
\end{eqnarray*}
where $y_{i,\obs}$ is the observed part of $(y_{1i},y_{2i})$. 
\item [{{{[}M-step{]}}}] Update the parameter $\theta$ by maximizing
$Q_{1}(\theta_{1}\mid\theta^{(t)})$ and $Q_{2}(\theta_{2}\mid\theta^{(t)})$
with respect to $\theta_{1}$ and $\theta_{2}$. 
\end{description}
The E-step and M-step can be iteratively computed until convergence,
leading to the pseudo maximum likelihood estimator $\widehat{\theta}$.

Given $\widehat{\theta}$, mass imputation can be done for both $Y_{2}$
in sample B and $Y_{1}$ in sample C. The imputation model for $Y_{2}$
in sample B is $f_{2}(Y_{2}\mid X,Y_{1};\widehat{\theta}_{2})$. Also,
the imputation model for $Y_{1}$ in sample C is 
\begin{equation}
f(Y_{1}\mid X,Y_{2};\widehat{\theta})=\frac{f_{1}(Y_{1}\mid X;\widehat{\theta}_{1})f_{2}(Y_{2}\mid X,Y_{1};\widehat{\theta}_{2})}{\int f_{1}(Y_{1}\mid X;\widehat{\theta}_{1})f_{2}(Y_{2}\mid X,Y_{1};\widehat{\theta}_{2})\de Y_{1}}.\label{5}
\end{equation}
To generate imputed values from (\ref{5}), one may use Markov Chain
Monte Carlo methods or the parametric fractional imputation of \citet{kim2011parametric}.

We now consider the non-monotone missingness II structure in Table
\ref{tab:Missingness-patterns}. Sample A and sample B are probability
samples were selected from the same finite population. In sample A,
observe $(X,Y_{1})$ and in sample B, observe $(X,Y_{2})$. The question
of interest is the associational relationship of $Y_{1}$ and $(X,Y_{2})$.
If $(X,Y_{1},Y_{2})$ were jointly observed, one can fit a simple
regression model of $Y_{2}$ on $(X,Y_{2}).$ However, based on the
available data, $Y_{1}$ and $Y_{2}$ not available simultaneously.

This problem fits into the statistical matching framework \citep{dorazio06}.
In statistical matching, the goal is to create $Y_{1}$ for each unit
in sample B by finding a ``statistical twin'' from the sample A.
Typically, one assumes the conditional independence assumption that
$Y_{1}$ and $Y_{2}$ are conditionally independent given $X,$ or
equivalently,

\begin{equation}
f(Y_{1}\mid X,Y_{2})=f(Y_{1}\mid X).\label{cia}
\end{equation}
Then, the ``statistical twin'' is solely determined by ``how close''
they are in terms of $X$'s. However, in a regression model of $Y_{1}$
on $(X,Y_{2})$, (\ref{cia}) sets the regression coefficient associated
with $Y_{2}$ to be zero \textit{a priori}, which is contrary to the
study question of interest.

For a joint modeling of $(X,Y_{1},Y_{2})$ without assuming (\ref{cia}),
identification is an important issue. Consider the following joint
model of $(Y_{1},Y_{2})$ given $X,$ 
\begin{eqnarray}
Y_{1} & = & \alpha_{0}+\alpha_{1}X+e_{1},\label{eq:ncia}\\
Y_{2} & = & \beta_{0}+\beta_{1}X+\beta_{2}Y_{1}+e_{2},\label{eq:ncia2}
\end{eqnarray}
where $\cov(e_{1},e_{2})=0$. Because $(X,Y_{1})$ is observed in
sample A, $(\alpha_{0},\alpha_{1})$ is identifiable. Because $(X,Y_{2})$
is observed in sample B, $f(Y_{2}\mid X)$ is identifiable.

Coupling (\ref{eq:ncia}) and (\ref{eq:ncia2}) leads to
\[
Y_{2}=(\beta_{0}+\alpha_{0}\beta_{2})+(\beta_{1}+\alpha_{1}\beta_{2})X+\beta_{2}e_{1}+e_{2}.
\]
Thus, only $\beta_{0}+\alpha_{0}\beta_{2}$ and $\beta_{1}+\alpha_{1}\beta_{2}$
are identifiable and $(\beta_{0},\beta_{1},\beta_{2})$ is not.

In general, adding non-linear terms in the parametric assumption can
help achieve identification. For example, consider adding $X^{2}$
in (\ref{eq:ncia}) such that 
\begin{eqnarray}
Y_{1} & = & \alpha_{0}+\alpha_{1}X+\alpha_{2}X^{2}+e_{1}.\label{eq:ncia3}
\end{eqnarray}
Again, $(\alpha_{0},\alpha_{1},\alpha_{2})$ is identifiable from
sample A. Coupling (\ref{eq:ncia2}) and (\ref{eq:ncia3}) leads to
\[
Y_{2}=(\beta_{0}+\alpha_{0}\beta_{2})+(\beta_{1}+\alpha_{1}\beta_{2})X+(\alpha_{2}\beta_{2})X^{2}+\beta_{2}e_{1}+e_{2}.
\]
Thus, $\beta_{0}+\alpha_{0}\beta_{2}$, $\beta_{1}+\alpha_{1}\beta_{2}$
and $\alpha_{2}\beta_{2}$ are identifiable from sample B. As long
as $\alpha_{2}\neq0$, $(\beta_{0},\beta_{1},\beta_{2})$ is then
identifiable. For an identifiable model, parameter estimation can
be implemented either using the EM algorithm or GLS.

Other assumptions can be invoked to achieve model identification.
\citet{kim2016} used an instrumental variable assumption for model
identification and develop fractional imputation methods \textcolor{black}{for
statistical matching}. \citet{park2016} presented an application
of the statistical matching technique using fractional imputation
in the context of handling mixed-mode surveys. \citet{park2017} applied
the method to combine two surveys with measurement errors.

\section{Combining probability and non-probability samples\label{sec:PNP}}

\subsection{Combining a probability sample with a non-probability sample}

 Statistical analysis of non-probability survey
samples faces many challenges as documented by \citet{baker13}. Non-probability
samples have unknown selection/inclusion mechanisms and are typically
biased, and they do not represent the target population. A popular
framework in dealing with the biased non-probability samples is to
assume that auxiliary variable information on the same population
is available from an existing probability survey sample. This framework
was first used by \citet{rivers2007} and followed by a number of
other authors including \citet{Vavreck2008}, \citet{Lee2009}, \citet{Valliant2011},
\citet{Brick2015}, \citet{elliott2017inference} and \citet{chen2018doubly},
among others. Combining the up-to-date information from a non-probability
sample and auxiliary information from a probability sample can be
viewed as data integration, which is an emerging area of research
in survey sampling \citep{lohr2017combining}. 

Data integration for finite population inference is similar to the
problem of combining randomized experiments and non-randomized real-world
evidence studies for causal inference of treatment effects \citep{keiding2016perils}.
In randomized clinical trial, the treatment assignment mechanism is
known and therefore treatment effect evaluation based on randomized
clinical trial is unconfounded. However, due to restrictive inclusion
and exclusion criteria, the trial sample may be narrowly defined and
can not represent the real-world patient population. On the other
hand, by the real-world data collection mechanism, the real-world
evidence study is often representative of the target population. Combining
trial and real-world evidence studies can achieve more robust and
efficient inference of treatment effect for a target patient population.
Table \ref{tab:Data-integration} draws a parallel comparison of data
sources between data integration in survey sampling and that in treatment
effect evaluation.

\begin{table}[H]
\caption{\label{tab:Data-integration}Data integration in survey sampling and
Biostatistics: }

\centering%
\begin{tabular}{cccc}
\hline 
 &  & Representative of & unbiased\tabularnewline
Survey sampling & Treatment effect evaluation & the finite population & estimation$^{*}$\tabularnewline
\hline 
Probability sample & Real world evidence study & $\checked$ & \tabularnewline
Non-probability sample & Randomized experiment &  & $\checked$\tabularnewline
\hline 
\end{tabular}

$^{*}$In survey sampling, some probability samples may not observe
the study variable of interest; for treatment effect evaluation, randomized
experiments provide unbiased estimation of treatment effect due to
treatment randomization.
\end{table}

Survey statisticians and biostatisticians have provided different
methods for combining information from multiple data sources. \citet{lohr2017combining}
and \citet{rao2019} provided comprehensive reviews of statistical
methods for finite population inference. Existing methods for data
integration of a probability sample and a non-probability sample can
be categorized into three types as follows. The first type is the
so-called propensity score adjustment \citep{rosenbaum1983central}.
In this approach, the probability of a unit being selected into the
non-probability sample, which is referred to as the propensity or
sampling score, is modeled and estimated for all units in the non-probability
sample. The subsequent adjustments, such as propensity score weighting
or stratification, can then be used to adjust for selection biases;
see, e.g., \citet{lee2009estimation,valliant2011estimating,elliott2017inference}
and \citet{chen2018doubly}. \citet{stuart2011use,stuart2015assessing}
and \citet{buchanan2018generalizing} used propensity score weighting
to generalize results from randomized trials to a target population.
\citet{o2014generalizing} proposed propensity score stratification
for analyzing a non-randomized social experiment. One notable disadvantage
of the propensity score methods is that they rely on an explicit propensity
score model and are biased and highly variable if the model is misspecified
\citep{kang2007demystifying}. The second type uses calibration weighting
\citep{deville1992calibration,kott2006using}. This technique calibrates
auxiliary information in the non-probability sample with that in the
probability sample, so that after calibration the weighted distribution
of the non-probability sample is similar to that of the target population
\citep{disogra2011calibrating}. The third type is mass imputation,
which imputes the missing values for all units in the probability
sample. In the usual imputation for missing data analysis, the respondents
in the sample constitute a training dataset for developing an imputation
model. In the mass imputation, an independent non-probability sample
is used as a training dataset, and imputation is applied to all units
in the probability sample; see, e.g., \citet{breidt1996two}; \citet{rivers2007};
\citet{kim2012combining}; \citet{chipperfield2012combining}; \citet{bethlehem2016solving};
and \citet{yang2018integration}.

\subsection{Setup and assumptions}

Non-probability samples become increasingly popular in survey statistics
but may suffer from selection bias that limits the generalizability
of results to the target population. We consider integrating a non-probability
sample with a carefully designed probability sample which provides
the representative covariate information of the target population.

Let $X\in\R^{p}$ be a vector of auxiliary variables (including an
intercept) that are available from two data sources, and let $Y\in\R$
be the study variable of interest. We consider combining a probability
sample with $X$, referred to as sample A, and a non-probability sample
with $(X,Y)$, referred to as sample B, to estimate $\mu_{y}$ the
population mean of $Y$. We focus on the case when the study variable
$Y$ is observed in \textcolor{black}{sample B} only, but the other
auxiliary variables are commonly observed in both data. Although the
big data source has a large sample size, the sampling mechanism is
often unknown, and we cannot compute the first-order inclusion probability
for Horvitz-Thompson estimation. The naive estimators without adjusting
for the sampling process are subject to selection biases, as illustrated
in Table \ref{tab:Illustration-of-bias-big-Data}. On the other hand,
although the probability sample with sampling weights represents the
finite population, it does not observe the study variable. The complementary
features of probability samples and non-probability samples raise
the question of whether it is possible to develop data integration
methods that leverage the advantages of both sources.

\begin{table}[H]
\caption{\label{tab:Illustration-of-bias-big-Data}Illustration of the total
error from the simple mean estimator of $\bar{Y}_{N}$ based on probability
simple random sample and big non-probability sample}

\centering%
\begin{tabular}{ccccc}
\hline 
Total Error (MSE)  & =  & Variance  & +  & Bias$^{2}$\tabularnewline
\hline 
Probability sample  &  & $\{(1-f_{A})/n_{A}\}S_{Y}^{2}$  &  & $0$\tabularnewline
Non-probability sample (Big data)  &  & $\approx0$  &  & $r_{B}^{2}\{(1-f_{B})/f_{B}\}S_{Y}^{2}$\tabularnewline
\hline 
\end{tabular}

$f_{A}=n_{A}/N$ and $f_{B}=n_{B}/N$ are the sampling fractions of
sample A and sample B, respectively; $r_{B}$ is the correlation between
the outcome $Y$ and the inclusion indicator $I_{B}$; $S_{Y}$ is
the population variance of $Y$. 
\end{table}

Because the sampling mechanism of a non-probability sample is unknown,
the target population quantity is not identifiable in general. Unlike
the previous case in Section \ref{sec:integProbSamples}, the sampling
mechanism of sample B is unknown and therefore $\mu_{y}$ is not identifiable
in general.

Two datasets were considered from the 2005 Pew Research Centre (PRC) and the 2005 Behavioral Risk Factor Surveillance
System (BRFSS). The goal of the PRC study was to evaluate the relationship
between individuals and community \citep{chen2018doubly,kim2018combining}.
The 2005 PRC data are non-probability sample data provided by eight
different vendors, which consist of $n_{B}=9,301$ subjects. \citet{yang2019doubly}
focus on two study variables, a continuous $Y_{1}$ (days had at least
one drink last month) and a binary $Y_{2}$ (an indicator of voted
local elections). The 2005 BRFSS sample is a probability sample, which
consists of $n_{A}=441,456$ subjects with survey weights. This dataset
does not have measurements on the study variables of interest; however,
it contains a rich set of common covariates with the PRC datase. The
covariate distributions from the PRC sample and the BRFSS sample are
considerably different, e.g., age, education (high school or less),
financial status (no money to see doctors, own house), retirement
rate, and health (smoking). Therefore, the PRC dataset is not representative
of the target population, and the naive analyses of the study variables
are subject to selection biases.

Let $f(Y\mid X)$ be the conditional distribution of $Y$ given $X$
in the superpopulation model $\zeta$ that generates the finite population.
We make the following assumption.

\begin{assumption} \label{asmp:MAR} (i) The sampling indicator $I_{B}$
of sample B and the \textcolor{black}{study} variable $Y$ is independent
given $X$; i.e. $P(I_{B}=1\mid X,Y)=P(I_{B}=1\mid X)$, referred
to as the sampling score $\pi_{B}(X)$; and (ii)\textcolor{black}{{}
$\pi_{B}(X)>0$ }for all $X$.

\end{assumption}

Assumption \ref{asmp:MAR} (i) and (ii) constitute the strong ignorability
condition \citep{rosenbaum1983central}. This assumption holds if
the set of covariates contains all predictors for the outcome that
affect the possibility of being selected in sample B. This setup has
previously been used by several authors; see, e.g., \citet{rivers2007}
and \citet{Vavreck2008}. Assumption \ref{asmp:MAR} (i) states the
ignorability of the selection mechanism to sample B conditional upon
the covariates. Under Assumption \ref{asmp:MAR} (i), $E(Y\mid X)=E(Y\mid X,I_{B}=1)$,
denoted by $m(X)$, can be estimated based on sample B. Assumption
\ref{asmp:MAR} (ii) implies that the support of in sample B is the
same as that in the finite population. Assumption \ref{asmp:MAR}
(ii) does not hold if certain units would never be included in the
\textcolor{black}{non-probability} sample. The plausibility of this
assumption can be \textcolor{black}{easily checked by comparing the
marginal distributions of the auxiliary variables in sample B with
those in sample A.}

Under the sampling ignorability assumption, there are two main approaches:
i) the weighting approach by constructing weights for sample B to
improve the representativeness of sample B; ii) the imputation approach
by creating mass imputation for sample A using the observations in
sample B. There is considerable interest in bridging the findings
from a randomized clinical trial to the target population. This problem
has been termed as generalizability \citep{cole2010generalizing,stuart2011use,hernan2011compound,tipton2013improving,o2014generalizing,stuart2015assessing,keiding2016perils,buchanan2018generalizing},
external validity \citep{rothwell2005external} or transportability
\citep{pearl2011transportability,rudolph2017robust} in the statistics
literature and has connections to the covariate shift problem in machine
learning \citep{sugiyama2012machine}.

\subsection{Propensity score weighting}

\textcolor{black}{{} Under Assumption (i) and (ii), we can build a model
for $\pi_{B}(X)=P(I_{B}=1\mid X)$ and use it to adjust for the selection
bias in sample B. } In practice, the propensity score function $\pi_{B}(X)$
is unknown and needs to be estimated from the data. Let $\pi_{B}(X;\alpha)$
be the posited models for $\pi_{B}(X)$, where $\alpha$ is the unknown
parameter. Several authors have proposed different estimation strategies.
For example, $\widehat{\alpha}$ can be obtained by a weighted regression
of $I_{B,i}$ on $x_{i}$ combining sample A and sample B ($I_{B,i}=0$
for $i\in A$ and $I_{B,i}=1$ for $i\in B$), weighted by the design
weights from sample A, which is valid if the size of sample B is relatively
small \citep{valliant2011estimating}. \citet{chen2018doubly} proposed
estimating $\alpha$ by solving 
\begin{equation}
\widehat{S}_{1}(\alpha)=\sum_{i\in B}x_{i}-\sum_{i\in A}d_{A,i}\pi_{B}(x_{i};\alpha)x_{i}=0,\label{eq:(5)}
\end{equation}
which is a sample version of the population estimating equation $S(\alpha)=\sum_{i\in U}\left\{ I_{B,i}-\pi(x_{i};\alpha)\right\} x_{i}=0.$
Instead of using (\ref{eq:(5)}), one can also use 
\[
\text{\ensuremath{\widehat{S}_{2}(\alpha)}}=\sum_{i\in B}\frac{1}{\pi_{B}(x_{i};\alpha)}x_{i}-\sum_{i\in A}d_{A,i}x_{i}=0,
\]
which is closely related to the calibration weighting approach for
nonresponse adjustment.

Given $\widehat{\alpha}$, the inverse probability of sampling weighting
estimator of $\mu_{y}$ is

\begin{equation}
\widehat{\mu}_{\ipw}=\widehat{\mu}_{\ipw}(\widehat{\alpha})=N^{-1}\sum_{i=1}^{N}\frac{I_{B,i}}{\pi_{B}(x_{i};\widehat{\alpha})}y_{i}.\label{eq:psw}
\end{equation}
Variance estimation of $\widehat{\mu}_{\ipw}$ can be obtained by
the standard M-estimation theory.

One of the notable disadvantages of the propensity score methods is
that they rely on an explicit propensity score model and are biased
if the model is mis-specified \citep{kang2007demystifying}. Moreover,
if the estimated propensity score is close to zero, $\widehat{\mu}_{\ipw}$
will be highly unstable.

\subsection{Calibration weighting}

The second weighting strategy is calibration weighting, or bench marking
weighting \citep{deville1992calibration,kott2006using}. This technique
can be used to calibrate auxiliary information in the non-probability
sample with that in the probability sample, so that after calibration
the non-probability sample is similar to the target population \citep{disogra2011calibrating}.

Instead of estimating the propensity score model and inverting the
propensity score to correct for the selection bias of the non-probability
sample, the calibration strategy estimates the weights directly. Toward
this end, we assign a weight $\omega_{B,i}$ to each unit $i$ in
the sample B so that 
\begin{equation}
\sum_{i\in B}\omega_{B,i}x_{i}=\sum_{i\in A}d_{A,i}x_{i}.\label{eq:calibration constraints}
\end{equation}
where $\sum_{i\in A}d_{A,i}x_{i}$ is a design-weighted estimate of
the population total of $X$ from the probability sample. Constraint
\eqref{eq:calibration constraints} is referred to as the covariate
balancing constraint \textcolor{black}{{} \citep{imai2014}}, and weights
$\mathcal{Q}_{B}=\{\omega_{B,i}:i\in B\}$ are the calibration weights.
The balancing constraint calibrates the covariate distribution of
the non-probability sample to the target population in terms of $X$.
Instead of calibrating each $X,$ one can calibrate model-based calibration
\citep{mcconville2017model,chen2018model,chen2019calibrating}. In
this approach, one can posit a parametric model for $E(Y\mid X)=m(X;\beta)$
and estimate the unknown parameter $\beta$ based on sample B. The
model-based calibration specifies the constraints for $\mathcal{Q}_{B}$
as

\begin{equation}
\sum_{i\in B}\omega_{B,i}m(x_{i};\widehat{\beta})=\sum_{i\in A}d_{A,i}m(x_{i};\widehat{\beta})\label{eq:model-cal-constraint}
\end{equation}

We estimate $\mathcal{Q}_{B}$ by solving the following optimization
problem: 
\begin{equation}
\underset{\mathcal{Q}_{B}}{\text{min}}\left\{ L(\mathcal{Q}_{B})=\sum_{i\in B}\omega_{B,i}\log\omega_{B,i}\right\} ,\label{eq:optQ}
\end{equation}
subject to $\omega_{B,i}\ge0,\;$ for all $i\in B$; $\sum_{i\in B}\omega_{B,i}=N$,
and the balancing constraint (\ref{eq:calibration constraints}) or
(\ref{eq:model-cal-constraint}).

The objective function in \eqref{eq:optQ} is the entropy of the calibration
weights; thus, minimizing this criteria ensures that the empirical
distribution of calibration weights are not too far away from the
uniform, such that it minimizes the variability due to heterogeneous
weights. This optimization problem can be solved using convex optimization
with Lagrange multiplier. Other objective functions, such as $L(\mathcal{Q}_{B})=\sum_{i\in B}\omega_{B,i}^{2}$,
can also be considered. This optimization problem can be solved using
convex optimization with Lagrange multiplier. By introducing Lagrange
multiplier $\lambda$, the objective function becomes 
\begin{equation}
L(\lambda,\mathcal{Q}_{B})=\sum_{i\in B}\omega_{B,i}\log\omega_{B,i}-\lambda^{\top}\left\{ \sum_{i\in B}\omega_{B,i}x_{i}-\sum_{i\in A}d_{A,i}x_{i}\right\} .\label{eq:lagrange_obj}
\end{equation}
Thus by minimizing \eqref{eq:lagrange_obj}, the estimated weights
are 
\[
\omega_{B,i}=\omega_{B}(x_{i};\widehat{\lambda})=\frac{N\exp\left(\widehat{\lambda}^{\top}x_{i}\right)}{\sum_{i\in B}\exp\left(\widehat{\lambda}^{\top}x_{i}\right)},
\]
and $\widehat{\lambda}$ solves the equation 
\begin{equation}
U(\lambda)=\sum_{i\in B}\exp\left(\lambda^{\top}x_{i}\right)\left\{ x_{i}-\textcolor{black}{ \frac{1}{N} }\sum_{i\in A}d_{A,i}x_{i}\right\} =0,\label{e:lam}
\end{equation}
which is the dual problem to the optimization problem (\ref{eq:optQ}).

The calibration weighting estimator is 
\begin{equation}
\widehat{\mu}_{\Cal}=\frac{1}{N}\sum_{i=1}^{N}\omega_{B,i}I_{B,i}y_{i}.\label{eq:cal-1}
\end{equation}
Variance estimation of $\widehat{\mu}_{\Cal}$ can be obtained by
the standard M-estimation theory by treating $\lambda$ as the nuisance
parameter and (\ref{e:lam}) as the corresponding estimating equation.

The justification for $\widehat{\mu}_{\Cal}$ subject to constraint
(\ref{eq:calibration constraints}) relies on the linearity of the
outcome model, i.e., $m(X)=X^{\T}\beta^{*}$ for some $\beta^{*}$,
or the linearity of the inverse probability of sampling weight, i.e.,
$\{\pi_{B}(X)\}^{-1}=X^{\T}\alpha^{*}$ for some $\alpha^{*}$ (\citealp{fuller2009sampling};
Theorem 5.1). The linearity conditions are unlikely to hold for non-continuous
variables. In these cases, $\widehat{\mu}_{\Cal}$ may be biased.
The justification for $\widehat{\mu}_{\Cal}$ subject to constraint
(\ref{eq:model-cal-constraint}) relies on a correct specification
of $m(X;\beta)$ in the data integration problem.

\textcolor{black}{{} \citet{chan2016} generalize this idea further
to develop a general calibration weighting method that satisfies the
covariance balancing property with increasing dimensions of the control
variables $m(x)$. \citet{zhao2019} developed a unified approach
of covariate balancing PS method using Tailored loss functions. The
regularization techniques using penalty terms into the loss function
can be naturally incorporated into the framework and machine learning
methods, such as boosting, can be used. The covariate balancing condition,
or calibration condition, in (\ref{eq:calibration constraints}),
can be relaxed. \citet{zubizarreta2015} relaxed the exact balancing
constraints to some tolerance level. \citet{wong2019} used the theory
of reproducing Kernel Hilbert space to develop an uniform approximate
balance for covariate functions. }


\subsection{Mass imputation approach}

The third type is mass imputation, where the imputed values are created
for the whole elements in the probability sample. In the usual imputation
for missing data analysis, the respondents in the sample provide a
training dataset for developing an imputation model. In the mass imputation,
an independent big data sample is used as a training dataset, and
imputation is applied to all units in the probability sample. While
the mass imputation idea for incorporating information from big data
is very natural, the literature on mass imputation itself is very
sparse.

In a parametric approach, let $m(X;\beta)$ be the posited model for
$m(X)$, where $\beta\in\mathcal{R}^{p}$ is the unknown parameter.
Under Assumption \ref{asmp:MAR}, $\widehat{\beta}$ can be obtained
by fitting the model to sample B. We assume that $\widehat{\beta}$
is the unique solution to 
\[
\widehat{U}(\beta)=\sum_{i\in B}\left\{ y_{i}-m(x_{i};\beta)\right\} h(x_{i};\beta)=0
\]
for some $p$-dimensional vector $h(x_{i};\beta)$. Thus, we use the
observations in sample B to obtain $\widehat{\beta}$ and use it to
construct $\widehat{y}_{i}=m(x_{i};\widehat{\beta})$ for all $i\in A$.

Under some regularity conditions, the mass imputation estimator 
\[
\widehat{\mu}_{\I}=\widehat{\mu}_{\I}(\widehat{\beta})=N^{-1}\sum_{i\in A}d_{A,i}m(x_{i};\widehat{\beta})
\]
satisfies $\widehat{\mu}_{\I}=\widehat{\mu}_{\I}(\beta_{0})+o_{P}(n_{B}^{-1/2})$
where 
\begin{eqnarray*}
\widehat{\mu}_{\I}(\beta) & = & N^{-1}\sum_{i\in A}d_{A,i}m(x_{i};\beta)+n_{B}^{-1}\sum_{i\in B}\left\{ y_{i}-m(x_{i};\beta)\right\} h(x_{i};\beta)^{\T}c^{*},\\
c^{*} & = & \left\{ n_{B}^{-1}\sum_{i\in B}\dot{m}(x_{i};\beta_{0})h^{\T}(x_{i};\beta_{0})\right\} ^{-1}\left\{ N^{-1}\sum_{i=1}^{N}\dot{m}(x_{i};\beta_{0})\right\} ,
\end{eqnarray*}
where $\beta_{0}$ is the true value of $\beta$ and $\dot{m}(x;\beta)=\partial m(x;\beta)/\partial\beta$.

Also, 
\[
E\{\widehat{\mu}_{\I}(\beta_{0})-\mu_{y}\}=0,
\]
and 
\begin{multline*}
\var\left\{ \widehat{\mu}_{\I}(\beta_{0})-\mu_{y}\right\} =\var\left\{ N^{-1}\sum_{i\in A}d_{A,i}m(x_{i};\beta_{0})-N^{-1}\sum_{i\in U}m(x_{i};\beta_{0})\right\} \\
+E\left[n_{B}^{-2}\sum_{i\in B}E\left(e_{i}^{2}\mid x_{i}\right)\left\{ h(x_{i};\beta_{0})^{\T}c^{*}\right\} ^{2}\right],
\end{multline*}
where $e_{i}=y_{i}-m(x_{i};\beta_{0})$. The justification for $\widehat{\mu}_{\I}$
relies on a correct specification of $m(X;\beta)$ and the consistency
of $\widehat{\beta}$. If $m(X;\beta)$ is misspecified or $\widehat{\beta}$
is inconsistent, $\widehat{\mu}_{\I}$ can be biased. For variance
estimation, either linearization method or bootstrap method can be
used. See \citet{kim2018combining} for more details.

\subsection{Doubly robust estimation}

To improve the robustness against model misspecification, one can
consider combining the weighting and imputation approaches \citep{kim2018sampling}.
The doubly robust estimator employs both the propensity score and
the outcome models, which is given by

\begin{equation}
\widehat{\mu}_{\dr}=\widehat{\mu}_{\dr}(\widehat{\alpha},\widehat{\beta})=N^{-1}\sum_{i=1}^{N}\left[\frac{I_{B,i}}{\pi_{B}(x_{i};\widehat{\alpha})}\{y_{i}-m(x_{i};\widehat{\beta})\}+I_{A,i}d_{A,i}m(x_{i};\widehat{\beta})\right].\label{eq:dr}
\end{equation}

The estimator $\widehat{\mu}_{\dr}$ is doubly robust in the sense
that it is consistent if either the propensity score model or the
outcome model is correctly specified, not necessarily both. Moreover,
it is locally efficient if both models are correctly specified \citep{bang2005doubly,cao2009improving}.
Let $\widehat{\mu}_{\HT}=N^{-1}\sum_{i\in A}d_{A,i}y_{i}$ be the
Horvitz--Thompson estimator that could be used if $y_{i}$ were observed
in sample A. We express $\widehat{\mu}_{\dr}-\widehat{\mu}_{\HT}=-\sum_{i\in A}d_{A,i}\widehat{e}_{i}+\sum_{i\in B}\{\pi_{B}(x_{i};\widehat{\alpha})\}^{-1}\widehat{e}_{i}$where
$\widehat{e}_{i}=y_{i}-\widehat{y}_{i}$. To show the double robustness
of $\widehat{\mu}_{\dr}$, we consider two scenarios. In the first
scenario, if $\pi_{B}(X;\alpha)$ is correctly specified, then 
\[
E\left(\widehat{\mu}_{\dr}-\widehat{\mu}_{\HT}\mid\F_{N}\right)\cong-\sum_{i\in A}d_{A,i}\widehat{e}_{i}+\sum_{i\in U}\widehat{e}_{i}
\]
which is design-unbiased of zero. In the second scenario, if $m(X;\beta)$
is correctly specified, then $E(\widehat{e}_{i})\cong0$. In both
cases, $\widehat{\mu}_{\dr}-\widehat{\mu}_{\HT}$ is unbiased of zero
and therefore $\widehat{\mu}_{\dr}$ is unbiased of $\mu_{y}$.

If either $\pi_{B}(X^{\T}\alpha)$ or $m(X^{\T}\beta)$ is correctly
specified, 
\[
n^{1/2}\left\{ \widehat{\mu}_{\dr}(\widehat{\alpha},\widehat{\beta})-\mu\right\} \rightarrow\N\left(0,V\right),
\]
as $n\rightarrow\infty$, where $V=\lim_{n\rightarrow\infty}(V_{1}+V_{2}),$
\begin{eqnarray*}
V_{1} & = & E\left\{ \frac{n}{N^{2}}\sum_{i=1}^{N}\sum_{j=1}^{N}(\pi_{A,ij}-\pi_{A,i}\pi_{A,j})\frac{m(x_{i}^{\T}\beta^{*})}{\pi_{A,i}}\frac{m(x_{j}^{\T}\beta^{*})}{\pi_{A,j}}\right\} ,\\
V_{2} & = & \frac{n}{N^{2}}\sum_{i=1}^{N}E\left[\left\{ \frac{I_{B,i}}{\pi_{B,i}(x_{i}^{\T}\alpha^{*})}-1\right\} ^{2}\left\{ y_{i}-m(x_{i}^{\T}\beta^{*})\right\} ^{2}\right].
\end{eqnarray*}

To estimate $V_{1}$, we can use the design-based variance estimator
applied to $m(X_{i}^{\T}\widehat{\beta})$ as 
\begin{equation}
\widehat{V}_{1}=\frac{n}{N^{2}}\sum_{i\in A}\sum_{j\in A}\frac{(\pi_{A,ij}-\pi_{A,i}\pi_{A,j})}{\pi_{A,ij}}\frac{m(X_{i}^{\T}\widehat{\beta})}{\pi_{A,i}}\frac{m(X_{j}^{\T}\widehat{\beta})}{\pi_{A,j}}.\label{eq:Vhat1}
\end{equation}
To estimate $V_{2},$ we further express $V_{2}$ as 
\begin{equation}
V_{2}=\frac{n}{N^{2}}\sum_{i=1}^{N}E\left[\left\{ \frac{I_{B,i}}{\pi_{B,i}(X_{i}^{\T}\alpha^{*})^{2}}-\frac{2I_{B,i}}{\pi_{B,i}(X_{i}^{\T}\alpha^{*})}\right\} \left\{ Y_{i}-m(X_{i}^{\T}\beta^{*})\right\} ^{2}+\left\{ Y_{i}-m(X_{i}^{\T}\beta^{*})\right\} ^{2}\right].\label{eq:V2-1}
\end{equation}
Let $\sigma^{2}(X_{i}^{\T}\beta^{*})=E\left[\left\{ Y_{i}-m(X_{i}^{\T}\beta^{*})\right\} ^{2}\right]$,
and let $\widehat{\sigma}^{2}(X_{i})$ be a consistent estimator of
$\sigma^{2}(X_{i}^{\T}\beta^{*})$. We can then estimate $V_{2}$
by 
\begin{equation}
\widehat{V}_{2}=\frac{n}{N^{2}}\sum_{i=1}^{N}\left[\left\{ \frac{I_{B,i}}{\pi_{B}(X_{i}^{\T}\widehat{\alpha})^{2}}-\frac{2I_{B,i}}{\pi_{B}(X_{i}^{\T}\widehat{\alpha})}\right\} \left\{ Y_{i}-m(X_{i}^{\T}\widehat{\beta})\right\} ^{2}+I_{A,i}d_{A,i}\widehat{\sigma}^{2}(X_{i})\right].\label{eq:Vhat2}
\end{equation}
By the law of large numbers, $\widehat{V}_{2}$ is consistent for
$V_{2}$ regardless of whether one of $\pi_{B,i}(X_{i}^{\T}\alpha)$
or $\pi_{B,i}(X_{i}^{\T}\beta)$ is misspecified, and therefore it
is doubly robust.

\section{Combining probability and big data\label{sec:PbigD}}

\subsection{Big data sample}

\textcolor{black}{To meet the new challenges in the probability sampling,
statistical offices face the increasing pressure to utilize convenient
but often uncontrolled big data sources, such as satellite information
\citep{mcroberts2010advances}, mobile sensor data \citep{palmer2013new},
and web survey panels \citep{tourangeau2013science}. \citet{couper2013sky},
\citet{citro2014multiple}, \citet{tam2015big}, and \citet{pfeffermann2015methodological}
articulated the promise of harnessing big data for official and survey
statistics but also raised  many issues regarding big data sources.
While such data sources provide timely data for a large number of
variables and population elements, they are non-probability samples
and often fail to represent the target population of interest because
of inherent selection biases. \citet{tam2018big} also covered some
ethical challenges of big data for official statisticians and discuss
some preliminary methods of correcting for selection bias in big data.}

Combining information from several sources to improve estimates for
population parameters is an important practical problem in survey
sampling. In the past decade, more and more auxiliary information
became available, including large administrative record datasets
and remote sensing data derived from satellite images. How to combine
such information with survey data to provide better estimates for
population parameters is a new challenge that survey statisticians
face today. \citet{tam2015big} presented an overview of some initiatives
of big data applications in official statistics of the Australian
Bureau of Statistics. Such big data are becoming increasingly popular
and they come from a variety of sources such as remote sensing data,
administrative data such as tax data, so on.

Suppose that there are two data sources, one from a probability sample,
referred to as sample A, and the other from a big data source, referred
to as sample B. Table \ref{tab:Data-structure-for-big} illustrates
the observed data structure.

\begin{table}[H]
\caption{\label{tab:Data-structure-for-big}Data structure for data integration
with big data}

\centering%
\begin{tabular}{ccccc}
\hline 
 &  & $d$  & $X$  & $Y$\tabularnewline
\hline 
Scenario 1  & Sample A  & $\checked$  & $\checked$  & $\checked$\tabularnewline
 & Sample B  &  & $\checked$  & \tabularnewline
\hline 
Scenario 2  & Sample A  & $\checked$  & $\checked$  & \tabularnewline
 & Sample B  &  & $\checked$  & $\checked$\tabularnewline
\hline 
\end{tabular}

Sample A is a probability sample, and Sample B is a big data sample,
which may not be representative of the population. 
\end{table}

\subsection{Scenario 1: leverage auxiliary information in big data to improve
efficiency}

In Scenario 1, the probability sample contains $Y$ observations.
Therefore, $\mu_{y}$ is identifiable and can be estimated by the
commonly-used estimator solely from sample A, denoted by $\widehat{\mu}_{A}$.
We can leverage the $X$ information in the big data sample to improve
the sample A estimator. We consider the case when additionally the
membership to the big data can be determined throughout the probability
sample. The key insight is that the subsample of units in sample A
with the big data membership constitutes a second-phase sample from
the big data sample, which acts as a new population. We calibrate
the information in the second-phase sample to be the same as the new
acting population. The calibration process in turn improves the accuracy
of the mass imputation estimator without specifying any model assumptions.
Let $h=(I_{B},1-I_{B},I_{B}X)$.

Following \citet{yang2018combining}, we can consider a class of estimators
satisfying 
\begin{equation}
n_{A}^{1/2}\left(\begin{array}{c}
\widehat{\mu}_{A}-\mu_{y}\\
\widehat{h}_{A}-\widehat{h}_{B}
\end{array}\right)\rightarrow\N\left\{ 0,\left(\begin{array}{cc}
V_{yy,A} & \Gamma^{\T}\\
\Gamma & V
\end{array}\right)\right\} ,\label{eq:asymp}
\end{equation}
in distribution, as $n_{A}\rightarrow\infty$, where $\widehat{h}_{A}=N^{-1}\sum_{i\in A}d_{A,i}h_{i}$
and $\widehat{h}_{B}=N^{-1}\sum_{i\in B}h_{i}$. Heuristically, if
(\ref{eq:asymp}) holds exactly rather than asymptotically, by the
multivariate normal theory, we have the following the conditional
distribution 
\[
n_{A}^{1/2}(\widehat{\mu}_{A}-\mu_{y})\mid n_{A}^{1/2}(\widehat{h}_{A}-\widehat{h}_{B})\sim\mathcal{N}\left\{ n_{A}^{1/2}\Gamma^{\T}V^{-1}(\widehat{h}_{A}-\widehat{h}_{B}),V_{yy,A}-\Gamma^{\T}V^{-1}\Gamma\right\} .
\]
Let $\widehat{V}_{yy,A},$ $\widehat{\Gamma}$ and $\widehat{V}$
be consistent estimators for $V_{yy,A},$ $\Gamma$ and $V$. We set
$n_{A}^{1/2}(\widehat{\mu}_{A}-\mu_{y})$ to equal its estimated conditional
mean $n_{A}^{1/2}\widehat{\Gamma}^{\T}\widehat{V}^{-1}(\widehat{h}_{A}-\widehat{h}_{B})$,
leading to an estimating equation for $\mu_{y}$: 
\[
n_{A}^{1/2}(\widehat{\mu}_{A}-\mu_{y})=n_{A}^{1/2}\widehat{\Gamma}^{\T}\widehat{V}^{-1}(\widehat{h}_{A}-\widehat{h}_{B}).
\]
Solving this equation for $\mu_{y}$, we obtain the estimator 
\begin{equation}
\widehat{\mu}=\widehat{\mu}_{A}-\widehat{\Gamma}^{\T}\widehat{V}^{-1}(\widehat{h}_{A}-\widehat{h}_{B}).\label{eq:proposed estimator}
\end{equation}

Under certain regularity conditions, if (\ref{eq:asymp}) holds, then
$\widehat{Y}$ is consistent for $\bar{Y}_{N}$, and 
\begin{equation}
n_{A}^{1/2}(\widehat{\mu}-\mu_{y})\rightarrow\N(0,V_{yy,A}-\Gamma^{\T}V^{-1}\Gamma),\label{eq:asymp var}
\end{equation}
in distribution, as $n_{A}\rightarrow\infty$. Given a nonzero $\Gamma$,
the asymptotic variance, $V_{yy,A}-\Gamma^{\T}V^{-1}\Gamma,$ is smaller
than the asymptotic variance of $\widehat{\mu}_{A}$, $V_{yy,A}$.

The asymptotic variance of $\widehat{\mu}$ can be estimated by 
\begin{equation}
\widehat{V}=(\widehat{V}_{yy,A}-\widehat{\Gamma}^{\T}\widehat{V}^{-1}\widehat{\Gamma})/n_{A}.\label{eq:VE}
\end{equation}

\citet{kimtam18} also explored similar ideas. They develop a calibration
weighting method to incorporate the big data auxiliary information
and apply the method to the official statistics in Australian Bureau
of Statistics. In this application, the big data is the Australian
Agricultural Census with 85\% response rate and the probability sample
is the Rural Environment and Agricultural Commodities Survey used
for calibration. In this application, the measurement from Census
data is the auxiliary variable used for calibration.

\subsection{Scenario 2: leverage probability sampling designs to correct for
selection bias}

In Scenario 2, we have a similar setup as in Section \ref{sec:PNP}.
Depending on the roles in statistical inference, there are two types
of \textit{big data}: one with large sample sizes (large $n$) and
the other with rich covariates (large $p$). We review methods for
the two types of big data.

\subsubsection{Robust mass imputation estimation}

In the first type, the non-probability sample can be large in sample
size. How to leverage the rich information in the big data to improve
the finite population inference is an important research. We review
robust mass imputation methods.

When the sample size of the big data is large, mass imputation is
more desirable. In mass imputation, we can train a predictive model
from the big data and impute the missing $y_{i}$ in sample A. Instead
of a parametric approach, we can also consider nonparametric approaches.
To find suitable imputed values, we consider nearest neighbor imputation;
that is, find the closest matching unit from sample B based on the
$X$ values and use the corresponding $Y$ value from this unit as
the imputed value.

Using sample B (big data) as a training data, find the nearest neighbor
of each unit $i\in A$ using a distance measure $d(x_{i},x_{j})$.
Let $i(1)$ be the index of its nearest neighbor, which satisfies
\[
d(x_{i(1)},x_{i})\leq d(x_{j},x_{i}),\forall j\in B.
\]
The nearest neighbor imputation estimator of $\mu$ is 
\[
\widehat{\mu}_{\nni}=N^{-1}\sum_{i\in A}d_{A,i}y_{i(1)}.
\]

\citet{yang2018integration} showed that under some regularity conditions,
$\widehat{\mu}_{\nni}$ has the same asymptotic distribution as $\widehat{\mu}_{\HT}=N^{-1}\sum_{i\in A}d_{A,i}y_{i}$.
Therefore, the variance of $\widehat{\mu}_{\nni}$ is the same as
the variance of $\widehat{\mu}_{\HT}.$ This implies that the standard
point estimator can be applied to the imputed data$\{(x_{i},y_{i(1)}):i\in A\}$
as if the $y_{i(1)}$'s were observed values. Let $\pi_{A,ij}$ be
the joint inclusion probability for units $i$ and $j$. They  showed that the direct variable estimator based
on the imputed data 
\[
\widehat{V}_{\nni}=\frac{n_{A}}{N^{2}}\sum_{i\in A}\sum_{j\in A}\frac{(\pi_{A,ij}-\pi_{A,i}\pi_{A,j})}{\pi_{A,ij}}\frac{y_{i(1)}}{\pi_{A,i}}\frac{y_{j(1)}}{\pi_{A,j}}
\]
is consistent for $V_{\nni}.$

\citet{yang2018integration}  also considered two strategies for improving the nearest neighbor
imputation estimator, one using $K$-nearest neighbor imputation \citep{mack1979multivariate}
and the other using generalized additive models \citep{wood2006generalized}.
In $K$-nearest neighbor imputation, instead of using one nearest
neighbor, they identify multiple nearest neighbors in the big data sample
and use the average response as the imputed value. This method is
popular in the international forest inventory community for combining
ground-based observations with imagines from remote sensors \citep{mcroberts2010advances}.
In the second strategy, they investigated modern techniques of prediction
for mass imputation with flexible models. They used generalized additive
models \citep{wood2006generalized} to learn the relationship of the
outcome and covariates from the big data and create predictions for
the probability samples. We note that this strategy can apply to a
wider class of semi- and non-parametric estimators such as single
index models, Lasso estimators \citep{belloni2015some}, and machine
learning methods such as random forests \citep{breiman1984classification}.

\subsubsection{Variable selection in the presence of a large number of covariates}

In the second type, when there are a large number of variables, there
is a large literature on variable selection methods for prediction,
but little work on variable selection for data integration that \textcolor{black}{can
successfully recognize the strengths and the limitations of each data
source and utilize all information captured for finite population
inference.}

In practice, subject matter experts recommend a rich set of potentially
useful variables but typically will not identify the set of variables
to adjust for. In the presence of a large number of auxiliary variables,
variable selection is important, because existing methods may become
unstable or even infeasible, and irrelevant auxiliary variables can
introduce a large variability in estimation. \citet{gao2017data}
proposed a pseudo-likelihood approach for combining multiple non-survey
data with high dimensionality; this approach requires all likelihoods
be correctly specified and therefore is sensitive to model misspecification.
\citet{chen2018model} proposed a model-based calibration approach
using LASSO; this approach relies on a correctly specified outcome
model.

\citet{yang2019doubly} proposed a doubly robust variable selection
and estimation strategy. In the first step, it selects a set of variables
that are important predictors of either the sampling score or the
outcome model using penalized estimating equations. In the second
step, it re-estimates the nuisance parameter $(\alpha,\beta)$ based
on the joint set of covariates selected from the first step and considers 
a doubly robust estimator of $\mu$, $\widehat{\mu}_{\dr}(\widehat{\alpha},\widehat{\beta})$
in (\ref{eq:dr}), where the estimating functions are 
\begin{equation}
J(\alpha,\beta)=\left(\begin{array}{c}
J_{1}(\alpha,\beta)\\
J_{2}(\alpha,\beta)
\end{array}\right)=\left(\begin{array}{c}
N^{-1}\sum_{i=1}^{N}I_{B,i}\left\{ \frac{1}{\pi_{B}(x_{i}^{\T}\alpha)}-1\right\} \{y_{i}-m(x_{i}^{\T}\beta)\}x_{i}\\
N^{-1}\sum_{i=1}^{N}\left\{ \frac{I_{B,i}}{\pi_{B}(x_{i}^{\T}\alpha)}-d_{A,i}I_{A,i}\right\} \partial m(x_{i}^{\T}\beta)/\partial\beta
\end{array}\right).\label{eq:jee}
\end{equation}

Importantly, the two-step estimator allows model misspecification
of either the sampling score or the outcome model. In the existing
high-dimensional causal inference literature, the doubly robust estimators
have been shown to be robust to selection errors using penalization
\citep{farrell2015robust} or approximation errors using machine learning
\citep{chernozhukov2018double}. However, this double robustness feature
requires both nuisance models to be correctly specified. Using (\ref{eq:jee})
relaxes this requirement by allowing one of the nuisance models to
be misspecified. This also enables one to construct a simple and consistent
variance estimator (\ref{eq:Vhat1})$+(\ref{eq:Vhat2})$ allowing
for doubly robust inferences.

\section{Concluding remarks\label{sec:Concluding-remark}}

Data integration is an emerging area of research with many potential
research topics. We have reviewed statistical techniques and applications
for data integration in survey sampling context. Probability sampling
remains as the gold standard to obtain a representative sample, but
the measurement of the study variable can be obtained from an independent
non-probability sample or big data. In this case, assumptions about
the sampling model or the outcome model are required. Most data integration
methods are based on the unverifiable assumption that the sampling
mechanism for the non-probability sample (or big data) is non-informative
(corresponding to the missingness at random in the missing data literature).

If the sampling mechanism is informative, imputation techniques can
be developed under the strong model assumptions for the sampling mechanism
\citep[e.g.,][]{riddles2016propensity,morikawa2018note}. Like the
non-informative sampling case, the informative sampling assumption
is unverifiable. In such settings, sensitivity analysis is recommended
to assess the robustness of the study conclusions to unverifiable
assumptions. This recommendation echoes Recommendation 15 of the National
Research Council (NRC) report entitled ``The Prevention and Treatment
of Missing Data in Clinical Trials'' \citep{NRC2010prevention}.
Chapter 5 of the NRC Report describes ``global'' sensitivity analysis
procedures that rigorously evaluate the robustness of study findings
to untestable assumptions about how missingness might be related to
the unobserved outcome.

When the training dataset has a hierarchical structure, multi-level
or hierarchical models can be used to develop mass imputation. This
is closely related to unit-level small area estimation in survey sampling
\citep{rao2015small}. The small area estimation is particularly promising
when we apply data integration using big data. \textcolor{black}{That
is, when we use big data as a training sample for prediction, the
multi-level model can be used to reflect the possible correlation
structure among observations. The parameter estimates for the multi-level
model computed from the big data can be used for predicting unobserved
study variables in the survey sample if the same multi-level model
can be made. Further research in this direction, including the mean
squared error estimation for this small area estimation, will be a
topic of future research.}

Finally, the uncertainty due to errors in record linkage and statistical matching is also an important problem. The matched sample using  recond linkage techniques \citep{fellegi1969} is subject to linakge errors. 
\cite{zhang2019} covers several research topics in the statistical analysis of  combined or fused data. 



\section*{Acknowledgment}
Dr. Yang is partially supported by the National Science Foundation grant DMS-1811245.  Dr. Kim is partially supported by the 
 National Science Foundation grant MMS-1733572 and  the Iowa Agriculture and Home Economics Experiment Station, Ames, Iowa. 

 \bibliographystyle{dcu}
\bibliography{DML,ci,Bibliography-MM-MC,pfi,bigdata,C:/Dropbox/bib/ci}

\end{document}